\documentclass[a4paper]{article}
\usepackage[mathscr]{eucal}
\usepackage{amsfonts,eucal,amssymb,latexsym,theorem}
\usepackage{fullpage}
\usepackage{graphicx}
\def\Par{\mathbin{\scalebox{-1}[-1]{\&}}}
\newtheorem{proposition}{Proposition}[section]
\newtheorem{theorem}{Theorem}[section]
{\theorembodyfont{\rmfamily}
\newtheorem{definition}{Definition}[section]}
\def\entails{\multimap}

\DeclareMathSymbol{\upPar}{\mathbin}{operators}{"26}
\def\subspace{\mathrel{\tilde{\subseteq}}}
\begin{document}
\title{Generalizing Topology via Chu Spaces\footnotemark}
\footnotetext{This is paper that was writen in 1999.}
\author{Basil K. Papadopoulos\\
         Department of Civil Engineering,\\
         Democritus University of Thrace,\\
         GR-671\ 00\ \ Xanthi, GREECE\\
         email: 
{\footnotesize {\upshape\texttt{papadob@civil.duth.gr}}} \and
        Apostolos Syropoulos\\
        Greek Molecular Computing Group\\
        366, 28th October Str.\\
        GR-g71\ 00\ \ Xanthi, Greece\\
{\footnotesize\upshape\texttt{asyropoulos@yahoo.com}}}
\maketitle 
\begin{abstract}
By using the representational power of Chu spaces we define the notion
of a generalized topological space (or GTS, for short), i.e., a mathematical 
structure that generalizes the notion of a topological space. We demonstrate
that these topological spaces have as special cases known topological spaces.
Furthermore, we develop the various topological notions and concepts for GTS. 
Moreover, since the logic of Chu spaces is linear logic, we give an
interpretation of most linear logic connectives as operators that 
yield topological spaces.
\end{abstract}  
\section{Introduction} 
Chu spaces are the objects of mathematics previously known as \textit{games}
\cite{games91},\thispagestyle{empty}
which are the result of a long evolution of the Chu construct, i.e., the
enrichment of category $\mathbf{Chu}(V,k)$ over the category $V$.
The Chu construct, which is named after Po-Hsiang Chu a student of
Michael Barr, appears in the theory of \hbox{*-autonomous} categories
(see \cite{barr79} for a detailed description of the various related
concepts and \cite{barr96} for a recent account). 
The theory of Chu spaces has been developed by Pratt 
in an effort to provide 
an alternative representation of types and processes \cite{Pr94}, and
formal languages and finite state automata \cite{Pr97b}. However, the 
representational power of Chu spaces is not limited to the above cases.
For example Lafont and Streicher in \cite{games91} report that vector spaces,
topologies and Girard's coherent spaces are among the mathematical entities 
that can be represented with Chu spaces; while Pratt in \cite{Pr95} reports 
that all ``partially distributive lattice'' categories can be realized by
them.  These remarks lead us to the definition of a generalization 
of the notion of a topological space which we call a
\textit{generalized topological space}, i.e., a topological space which, 
under certain conditions, has as special cases known topological spaces. 
We define all the usual notions associated with 
topological spaces, such as compactness, etc. Furthermore, since linear
logic \cite{Girard94lla} is the logic of Chu spaces, we give an
interpretation of most connectives of the logic as operators that yield
new topological spaces.
\section{Generalized Topological Spaces} %
We consider triplets of the form $(X,r,\mathfrak{A})$,
where $X$ is an arbitrary set called the \textit{set of points}, 
$\mathfrak{A}$ is an arbitrary set called the \textit{set of open sets}, 
and $r:X\times\mathfrak{A}\rightarrow I$ ($I=[0,1]$) is an arbitrary 
binary function called the \textit{membership function}. The intuitive meaning
of this function is that for all $x\in X$ and $A\in\mathfrak{A}$,
$r(x,A)$ is the degree to which $x$ is a member of the open set $A$.
Any such triplet will be called a \textit{generalized topological space}
(or GTS, for short).
\begin{definition}\label{GTS:def}
A GTS $(X,r,\mathfrak{A})$ is called a
\textit{strong generalized topological space} (or SGTS, for short)
if and only if the following two conditions are fulfilled:
\begin{enumerate}
\item If $A_{1},A_{2}\in\mathfrak{A}$, then there is an $A\in \mathfrak{A}$,
      such that
      \begin{displaymath} 
      \min\Bigl\{r(x,A_{1}),r(x,A_{2})\Bigr\}=r(x,A),
      \end{displaymath}
       for all 
      $x\in X$; we say that $A$ is the \textit{intersection}
      of $A_1$ and $A_2$, denoted as $A_1\cap A_2$. 
\item If $A_{j}\in\mathfrak{A}$, $j\in J$, there is an $A\in\mathfrak{A}$,
      such that $$\sup\Bigl\{r(x,A_{j}),j\in J\Bigr\}=r(x,A),$$ for all 
      $x\in X$; we say that $A$ is the \textit{infinite union} of $A_j$,
      denoted as $\bigcup_{j\in J}A_{j}$.
\end{enumerate}
\end{definition}
Note that we don't need to explicitly specify the elements of the union and
the intersection.
By imposing some restrictions on the range of values of $r$ and/or the
structure of $\mathfrak{A}$ we get common topological spaces: 
\paragraph{Fuzzy Topology} 
We assume that for a given SGTS, $\mathfrak{A}\subseteq I^{X}$, such that
$\bar{\mathbf{0}},\bar{\mathbf{1}}\in\mathfrak{A}$ 
($\bar{\mathbf{0}}(x)=0$ and $\bar{\mathbf{1}}(x)=1$ for all $x\in X$)
and so $r(x,A)=A(x)$, for
all $A\in\mathfrak{A}$ and all $x\in X$. Then the two conditions become
\begin{enumerate}
\item if $A_{1},A_{2}\in\mathfrak{A}$, then $A_{1}\wedge A_{2}\in\mathfrak{A}$,
\item if $\{A_{j}, j\in J\}\subseteq\mathfrak{A}$, then there is an 
$A\in\mathfrak{A}$, such that $\bigvee_{j\in J}A_{j}\in A$,
\end{enumerate} 
Then this is just the definition of a fuzzy topological space.
We proceed now with the definition of some useful concepts.
\begin{definition}
Let $(X,r,\mathfrak{A})$ be a GTS and $A_1,A_2\in\mathfrak{A}$, then
$A_1=A_2$ iff $r(x,A_1)=r(x,A_2)$ for all $x\in X$. Furthermore, we consider
two GTS $(X,r,\mathfrak{A})$ and $(X,s,\mathfrak{B})$, any $A\in\mathfrak{A}$
and $B\in\mathfrak{B}$ are equal iff $r(x,A)=s(x,B)$, for all $x\in X$.
\end{definition}
\begin{definition} Let $\mathcal{A}=(X,r,\mathfrak{A})$ and 
$\mathcal{B}=(Y,s,\mathfrak{B})$ be two GTSs, Then $\mathcal{B}$ is a
\textit{subspace} of $\mathcal{A}$, denoted as $\mathcal{B}\subspace
\mathcal{A}$, iff $Y\subseteq X$ and there is a surjection $\nu:\mathfrak{A}
\rightarrow\mathfrak{B}$ such that $r(y,A)=s(y,\nu(A))$ 
for all $y\in Y$ and for all $A\in\mathfrak{A}$.
\end{definition}
\subsection{Closed Generalized Topological Spaces} 
Let $\mathcal{A}=(X,r,\mathfrak{A})$ be a GTS and
$\mathfrak{K}$ be an arbitrary set, such that there is a unique bijection
$\varphi:\mathfrak{A}\rightarrow\mathfrak{K}$. Moreover, define a function 
$\bar{r}: X\times\mathfrak{K}\rightarrow I$, such that for all 
$K\in\mathfrak{K}$ 
\begin{displaymath}
\bar{r}(x,K)=1-r(x,\varphi^{-1}(K))
\end{displaymath}
Then the GTS 
$\bar{\mathcal{A}}=(X,\bar{r},\mathfrak{K})$ is called a \textit{closed
generalized topological space} (or CGTS, for short). We call $\mathfrak{K}$
the \textit{set of closed sets}. Obviously, $r(x,A)=1-\bar{r}(x,\varphi(A))$.
In case that $\mathfrak{A}\subseteq\mathcal{P}(X)$ for each $A\in\mathfrak{A}$,
$\varphi(A)=A^{\complement}$ (i.e., $\varphi(A)$ is the \textit{complement}
of $A$); and for each 
$K\in\mathfrak{K}$, $\varphi^{-1}(K)=K^{\complement}$. In case that 
$\mathfrak{A}\subseteq I^{X}$, then from the definition of $\bar{r}$,
we get that for all $K\in\mathfrak{K}$
\begin{eqnarray*}
K(x)         &=& \bar{r}(x,K)\\  
             &=& 1 - r(x,\varphi^{-1}(K))\\
             &=& 1 - r(x, K^{\complement})\\
             &=& 1 - K^{\complement}(x)
\end{eqnarray*}
Which is just the definition of the complement in fuzzy topology. The
following theorem proves that a SGTS and a bijection $\varphi$ induce
a CGTS with similar properties: 
\begin{theorem}
Let $\mathcal{A}=(X,r,\mathfrak{A})$ be a SGTS and $\varphi:\mathfrak{A}
\rightarrow\mathfrak{K}$ be a bijections such that
$\bar{\mathcal{A}}=(X,\bar{r},\mathfrak{K})$ is a CGTS. Then,
\begin{enumerate}
\item
if $K_1,K_2\in\mathfrak{K}$, there is a $K\in\mathfrak{K}$ such that
\begin{displaymath}
\max\Bigl\{\bar{r}(x,K_1),\bar{r}(x,K_2)\Bigr\}=\bar{r}(x,K),\quad
\forall x\in X
\end{displaymath}
and
\item
if $K_j\in\mathfrak{K}$ and $j\in J$, there is a $K\in\mathfrak{K}$ such that
\begin{displaymath}
\inf\Bigl\{\bar{r}(x,K_j),j\in J\Bigr\}=\bar{r}(x,K), \quad\forall x\in X
\end{displaymath}
\end{enumerate}
\end{theorem} 
Any CGTS which satisfies the two conditions of the previous theorem is 
called a \textit{strong closed generalized topological space}
(or SCGTS, for short).
\begin{proposition}
Let $\mathcal{A}=(X,r,\mathfrak{A})$ be a GTS and $\varphi_1:\mathfrak{A}
\rightarrow\mathfrak{K}$ be a bijection which induces the CGTS
$\bar{\mathcal{A}}=(X,\bar{r},\mathfrak{K})$. Moreover, 
let $\mathcal{B}=(Y,s,\mathfrak{B})$ be a subspace of $\mathcal{A}$
(i.e., among others there is a surjection 
$\nu_1:\mathfrak{A}\rightarrow\mathfrak{B}$), and
$\varphi_2:\mathfrak{B}\rightarrow\mathfrak{L}$ be a bijection such that
$\bar{\mathcal{B}}=(Y,\bar{s},\mathfrak{L})$ is a CGTS. 
Then, $\bar{\mathcal{B}}\subspace\bar{\mathcal{A}}$.  
\end{proposition}
\subsection{Dual Generalized Topological Spaces}  
The dual of a GTS $\mathcal{A}=(X,r,\mathfrak{A})$, denoted as
$\mathcal{A}^\bot$, is defined to be the triplet
$(\mathfrak{A},r\!\breve{\phantom{p}},X)$, where 
$r\!\breve{\phantom{p}}(A,x)$ denotes the degree to which the
open set $A\in\mathfrak{A}$ \textit{contains} the point $x$, i.e., 
$r\!\breve{\phantom{p}}(A,x)=r(x,A)$. This means 
that the dual of a GTS and the GTS itself externally behave in the same way.
Their only difference is their internal structure.
\subsection{Subset-hood} 
Let $A_{1},A_{2}\in\mathfrak{A}$ be two open sets of a GTS
 $\mathcal{A}=(X,r,\mathfrak{A})$, and let $K_{1},K_{2}\in\mathfrak{K}$ 
be two closed sets of $\bar{\mathcal{A}}$, then we say that
\begin{itemize}
\item $A_{1}$ is a subset of $A_{2}$ (denoted $A_{1}\subseteq A_{2}$)
iff $r(x,A_{1})\le r(x,A_{2})$ for all $x\in X$, 
\item $K_{1}\subseteq K_{2}$  iff 
$\bar{r}(x,K_{1})\le \bar{r}(x,K_{2})$ for all $x\in X$, 
\item $A$ is a subset of of $K$ iff $r(x,A)\le \bar{r}(x,K)$, for all $x\in X$,
and
\item $K$ is a subset of of $A$ iff $\bar{r}(x,K)\le r(x,A)$, for all $x\in X$.
\end{itemize}
\begin{proposition}
Consider a GTS $\mathcal{A}=(X,r,\mathfrak{A})$, a bijection 
$\varphi:\mathfrak{A}\rightarrow\mathfrak{K}$, which indices the CGTS
$\bar{\mathcal{A}}$, then
for any $A_{1},A_{2}\in\mathfrak{A}$ such that $\varphi(K_{1})=A_1$ and
$\varphi(K_{2})=A_2$:
\begin{eqnarray*}
A_{1}\subseteq A_{2} &\Leftrightarrow& 
         \varphi(A_{1})\supseteq\varphi(A_{2})\\ 
K_{1}\subseteq K_{2} &\Leftrightarrow& 
         \varphi^{-1}(K_{1})\supseteq\varphi^{-1}(K_{2}) 
\end{eqnarray*}
\end{proposition}
\subsection{Continuous Functions and Isomorphic Spaces} 
Continuous functions between two GTSs behave exactly like the morphisms of
any category $\mathbf{Chu}(V,k)$, i.e.,
we consider two arbitrary GTS $\mathcal{A}=(X,r,\mathfrak{A})$ and
$\mathcal{B}=(Y,s,\mathfrak{B})$; then a \textit{continuous function}
from $\mathcal{A}$ to $\mathcal{B}$ is a pair or functions $(f,\bar{f})$,
where $f:X\rightarrow Y$ and 
$\bar{f}:\mathfrak{B}\rightarrow\mathfrak{A}$, such that for all $x\in X$
and $B\in\mathfrak{B}$ the following equation is true:
\begin{displaymath}
s(f(x),B)=r(x,\bar{f}(B))
\end{displaymath}
Suppose that each function of the pair $(f,\bar{f})$ is a bijection, then we
call the spaces $\mathcal{A}$ and $\mathcal{B}$ \textit{isomorphic} to each
other and we denote this by $\mathcal{A}\cong\mathcal{B}$. Moreover, this
pair pair of functions is called an \textit{isomorphism}.
\begin{proposition}
Let $\mathcal{A}=(X,r,\mathfrak{A})$ and $\mathcal{B}=(Y,s,\mathfrak{B})$
be two different GTSs. Moreover, the bijections 
$\varphi_1:\mathfrak{A}\rightarrow\mathfrak{K}$
and $\varphi_2:\mathfrak{B}\rightarrow\mathfrak{L}$ induce the CGTS
$\bar{\mathcal{A}}=(X,\bar{r},\mathfrak{K})$ and 
$\bar{\mathcal{B}}=(Y,\bar{s},\mathfrak{L})$.
Then the continuous transformation $(f,\bar{f})$ 
from $\mathcal{A}$ to $\mathcal{B}$, induces the continuous
transformation $(f,\bar{f}^{\ast})$ from $\bar{\mathcal{A}}$ to
$\bar{\mathcal{B}}$, where $\bar{f}^{\ast}=\varphi_1\circ\bar{f}\circ
\varphi_2^{-1}$.
\end{proposition}
The following result is a direct consequence of the previous proposition:
\begin{proposition}
Let $\mathcal{A}=(X,r,\mathfrak{A})$ be a GTS and 
$\varphi_1:\mathfrak{A}\rightarrow\mathfrak{K}$ 
be a bijection such that $\bar{\mathcal{A}}=(X,\bar{r},\mathfrak{K})$
is a CGTS. Similarly, let $\mathcal{B}=(X,r,\mathfrak{A})$ be a GTS,
$\varphi_2:\mathfrak{B}\rightarrow\mathfrak{L}$  be a bijection such that
$\bar{\mathcal{B}}=(Y,\bar{s},\mathfrak{L})$. Then, 
\begin{displaymath}
\bar{s}(f(x),L)=\bar{r}(x,\bar{f}^{\ast}(L)), \quad\forall x\in X, 
\forall L\in\mathfrak{L}
\end{displaymath}
\end{proposition}
\section{Compact Spaces} 
Since, \textit{compactness} is a very useful topological concept we 
must provide a definition of it for our generalized topological spaces.
\begin{definition}
A GTS $\mathcal{A}=(X,r,\mathfrak{A})$ is \textit{compact} iff for every
family of open sets of $\mathfrak{A}$, i.e, $\{A_i,i\in I\}$, such that
$\sup\{r(x,A_i),i\in I\}>0$ for all $x\in X$, there is a finite subfamily, 
i.e., $\{A_j,j\in J\}$, where $J$ is a finite subset of $I$, such that 
$\sup\{r(x,A_j),j\in J\}>0$ for all $x\in X$. 
\end{definition}
Since, compactness is a property that must be preserved between
isomorphic GTS, we prove the following proposition:
\begin{proposition}
Let $\mathcal{A}=(X,r,\mathfrak{A})$ and 
$\mathcal{B}=(Y,s,\mathfrak{B})$ be two different GTSs. If
$\mathcal{A}$ is compact and $\mathcal{A}\cong\mathcal{B}$, then
$\mathcal{B}$ is also compact.
\end{proposition}
\section{Connected Spaces} 
We define the GTS $\mathbf{2}=(2,\phi,\mathcal{P}(2))$, where $2=\{0,1\}$
and $\phi(1,\{0\})=0$, $\phi(1,\{1\})=1$, etc.
\begin{definition}
Let $\mathcal{A}=(X,r,\mathfrak{A})$ be a GTS. Then, $\mathcal{A}$ is
called \textit{connected} iff there is no continuous function 
$g:X\rightarrow 2$ which is surjective.
\end{definition}
It is trivial to prove the following proposition:
\begin{proposition}
Let $\mathcal{A}$ and $\mathcal{B}$ be two GTS such that 
$\mathcal{A}\cong\mathcal{B}$. If $\mathcal{A}$ is connected, then
$\mathcal{B}$ is also connected.
\end{proposition} 
\section{Linear Implication---Pointwise Topologies}\label{limp} 
Let $\mathcal{A}=(X,r,\mathfrak{A})$ and $\mathcal{B}=(Y,s,\mathfrak{B})$
be two GTS, then we define the \textit{pointwise topological} space, denoted as
$\mathcal{A}\entails\mathcal{B}$, to be the GTS 
$(\mathcal{B}^{\mathcal{A}},t,X\times\mathfrak{B})$, with 
$t(f,(x,B))=s(f(x),B)$. A pointwise topology is nothing else than the
function space of \cite{Pr97b}. We now investigate the meaning of this
topology in the classical and the fuzzy cases.

In the classical case the equation $t(f,(x,B))=s(f(x),B)$ means that
$f(x)\in B$ iff $f\in\langle x,B\rangle$, where $\langle x,B\rangle=\{g\in
Y^X: g(x)\in B\}$, and  the set
$\{(x,B):x\in X,\; B\in \mathcal{O}(Y)\}$ form a subbase in the
function space $\mathscr{F}(X,Y)$. In the fuzzy setting we assume that
$\langle x,B\rangle:\mathscr{F}(X,Y)\rightarrow I$, so $\langle x,B\rangle(f)=
t(f,(x,B))=B(f(x))$. This equation defines a topology which we call
\textit{fuzzy pointwise topology} on the set $\mathscr{F}(X,Y)$.

Consider the GTS $\mathcal{A}\entails\mathcal{B}=
(\mathcal{B}^{\mathcal{A}},t,X\times\mathfrak{B})$ and the bijections
$\varphi_2:\mathfrak{B}\rightarrow\mathfrak{L}$ and
$\varphi_{\ast}=\mathrm{id}_{X}\times\varphi_{2}$, which induce the
CGTS $\bar{\mathcal{B}}$ and the CGTS:
\begin{displaymath}
\overline{\mathcal{A}\entails\mathcal{B}}=
(\mathcal{B}^{\mathcal{A}},\bar{t},X\times\mathfrak{L}),
\end{displaymath}
respectively. It is easy to prove that $\bar{t}(f,(x,L))=\bar{s}(f(x),L)$ 
for all $x\in X$, $L\in\mathfrak{L}$, and $f\in\mathcal{B}^{\mathcal{A}}$:
\begin{eqnarray*}
\bar{t}(f,(x,L)) &=& 1 - t(f,\varphi^{-1}_{\ast}(x,L))\\
                 &=& 1 - s(f(x),B)\\
                 &=& \bar{s}(f(x),\varphi_{2}(B))\\
                 &=& \bar{s}(f(x),L)
\end{eqnarray*}
Moreover, if $\varphi_1:\mathfrak{A}\rightarrow
\mathfrak{K}$ is a bijection, which induces the CGTS $\bar{\mathcal{A}}$,
then the GTS $\bar{\mathcal{A}}\entails\bar{\mathcal{B}}$ is defined to
be the triplet: $(\bar{\mathcal{B}}^{\bar{\mathcal{A}}},\tau,
X\times\mathfrak{L})$, such that $\tau(f,(x,L))=\bar{s}(f(x),L)$, for all
$x\in X$ and $L\in\mathfrak{L}$.
Furthermore, it is trivial to prove that $\overline{\mathcal{A}\entails
\mathcal{B}}\cong\bar{\mathcal{A}}\entails\bar{\mathcal{B}}$.
\section{Hausdorf Spaces and Regular Spaces} 
\begin{definition}
A GTS $\mathcal{A}=(X,r,\mathfrak{A})$ is called a \textit{Hausdorf} space 
iff for every $x_1,x_2\in X$, such that $x_1\not= x_2$, there are
$A_1,A_2\in\mathfrak{A}$, such that $r(x_1,A_1)>0$, $r(x_2,A_2)>0$,  and
$$\min\Bigl\{r(x,A_1),r(x,A_2)\Bigr\}=0, \quad\forall x\in X.$$
\end{definition}
It is trivial to prove that continuous functions preserve Hausdorf spaces.
Moreover, the following result can be proved easily :
\begin{proposition} Let $\mathcal{A}=(X,r,\mathfrak{A})$ be a GTS and
$\mathcal{B}=(Y,s,\mathfrak{B})$ be a Hausdorf space, then the pointwise
topological space $\mathcal{A}\entails\mathcal{B}$ is a Hausdorf space.
\end{proposition}
\begin{definition}
Let $\mathcal{A}=(X,r,\mathfrak{A})$ be a GTS and 
$\bar{\mathcal{A}}=(X,\bar{r},\mathfrak{K})$ be a CGTS such that
$\varphi:\mathfrak{A}\rightarrow\mathfrak{K}$ is a bijection. Then, we
call $\mathcal{A}$ a \textit{regular} space iff for all $x\in X$ and
all $K\in\mathfrak{K}$ there are $A_1,A_2\in\mathfrak{A}$ such that
$r(x,A_1)>0$, $K\subseteq A_2$, i.e., $\bar{r}(x',K)\le r(x',A_2)$ for
all $x'\in X$, and the following holds
\begin{displaymath}
\min\Bigl\{r(x',A_1),r(x',A_2)\Bigr\}=0, \quad\forall x'\in X
\end{displaymath} 
\end{definition} 
Regularity is preserved by isomorphisms:
\begin{proposition}
Let $\mathcal{A}=(X,r,\mathfrak{A})$ be a regular space and 
$\mathcal{B}=(Y,s,\mathfrak{B})$ a GTS, such that 
$\mathcal{A}\cong\mathcal{B}$, then $\mathcal{B}$ is also regular.
\end{proposition}
\begin{proposition} Let $\mathcal{A}=(X,r,\mathfrak{A})$ and
$\mathcal{B}=(Y,s,\mathfrak{B})$ be two GTSs. If $\mathcal{B}$ is
regular, then $\mathcal{A}\entails\mathcal{B}$ is also regular.
\end{proposition}
\section{The Tensor Product}  
We consider two GTSs $\mathcal{A}=(X,r,\mathfrak{A})$ and 
$\mathcal{B}=(Y,s,\mathfrak{B})$, then the \textit{tensor product}, denoted
as $\mathcal{A}\otimes\mathcal{B}$, is defined as follows:
\begin{eqnarray*}
\mathcal{A}\otimes\mathcal{B} &=& 
         (\mathcal{A}\entails\mathcal{B}^{\bot})^{\bot}\\
                              &=&
         (X\times Y, t\!\breve{\phantom{t}}, 
         (\mathcal{B}^{\bot})^{\mathcal{A}})
\end{eqnarray*}
where $t\!\breve{\phantom{t}}((x,y),f)=s\!\breve{\phantom{s}}(f(x),y)$ and
$f:X\rightarrow\mathfrak{B}$.
The topological space $\mathcal{A}\otimes\mathcal{B}$ has as set of points 
the set $X\times Y$ and as set of open sets all the \textit{induced} 
open sets $f$, i.e., $t\!\breve{\phantom{t}}((x,y),f)$ denotes the degree
to which $(x,y)$ belongs to the \textit{open set} $f$. 
This new GTS has a concrete interpretation in both 
ordinary and fuzzy topology.
As usual we start with the interpretation in ordinary topology.

Every function $f:X\rightarrow\mathcal{O}(Y)$ induces on the Cartesian
product $X\times Y$ a topology which has as subbase the set
$\mathscr{Y}=\{\langle f\rangle:\; f:X\rightarrow\mathcal{O}(Y)\}$,
where $\langle f\rangle=\{(x,y): y\in f(x)\}$. 
In the fuzzy case, every function $f:X\rightarrow\mathcal{F}(Y)$, induces
a fuzzy subset of $X\times Y$, denotes as $\langle f\rangle$, 
which is defined as follows:
\begin{displaymath}
\langle f\rangle(x,y)=\Bigl(f(x)\Bigr)(y)
\end{displaymath}
The set $\mathscr{Y}=\{\langle f\rangle:\; f:X\rightarrow \mathcal{F}(Y)\}$ 
is a subbase for a fuzzy topology on the product $X\times Y$.
\section{The Tensor Sum} 
Again, we consider two GTSs $\mathcal{A}=(X,r,\mathfrak{A})$ and
$\mathcal{B}=(Y,s,\mathfrak{B})$, then the \textit{tensor sum}
is defined as follows:
\begin{eqnarray*}
\mathcal{A}\Par\mathcal{B} &=& \mathcal{A}^{\bot}\entails\mathcal{B}\\
                           &=& (\mathcal{B}^{\mathcal{A}^{\bot}},
                                 \tau, \mathfrak{A}\times\mathfrak{B})
\end{eqnarray*}
where $\tau(f,(A,B))=s(f(A),B)$ and $f:\mathfrak{A}\rightarrow Y$.
Naturally, this defines a pointwise topology. In particular, in the case
of ordinary topologies we have that $f:\mathcal{O}(X)\rightarrow Y$
and $f\in \langle A,B\rangle$ iff $f(A)\in B$. Similar conclusions
can be derived for the fuzzy case (see section~\ref{limp}).
\section{Topological Sum and Product} %
We define the \textit{topological sum} and the \textit{topological product}
of any two GTS. Since, their definition make use of the concept of
the \textit{direct sum}, or just sum, of two sets $A$ and $B$,
denoted as $A+B$, we must say that $A+B=\{0\}\times A\cup\{1\}\times B$.
\begin{definition}
The topological sum of two GTS $\mathcal{A}=(X,r,\mathfrak{A})$ and
$\mathcal{B}=(Y,s,\mathfrak{B})$, denoted as $\mathcal{A}\oplus\mathcal{B}$,
is the triplet $(X+Y,t,\mathfrak{A}\times\mathfrak{B})$, where
$t((x,y),(0,A))=r(x,A)$ and $t((x,y),(1,B))=s(y,B)$.
\end{definition}
There is also a special GTS, $\mathsf{0}=(\emptyset,\alpha,\{1\})$,
with the property that for any triplet $\mathcal{A}=(X,r,\mathfrak{A})$
the following relations hold:
\begin{displaymath}
\mathcal{A}\oplus\mathsf{0}\cong\mathcal{A}\cong\mathsf{0}\oplus\mathcal{A}
\end{displaymath}
\begin{definition}
The topological product of two GTS $\mathcal{A}=(X,r,\mathfrak{A})$ and
$\mathcal{B}=(Y,s,\mathfrak{B})$, denoted as $\mathcal{A}\upPar\mathcal{B}$,
is the triplet $(X\times Y,t',\mathfrak{A}+\mathfrak{B})$, where
$t((0,x),(A,B))=r(x,A)$ and $t((1,y),(A,B))=s(y,B)$.
\end{definition}
The special GTS, $\top=(\{1\},\alpha\!\breve{\phantom{\alpha}},\emptyset)$,
has the property that for any triplet $\mathcal{A}=(X,r,\mathfrak{A})$
the following relations hold:
\begin{displaymath}
\mathcal{A}\upPar\top\cong\mathcal{A}\cong\top\upPar\mathcal{A}
\end{displaymath}
It is easy to verify that $\top=\mathsf{0}^{\bot}$ and to prove that 
$\mathcal{A}\oplus\mathcal{B}\cong(\mathcal{A}^{\bot}\upPar\mathcal{B}^{%
\bot})^\bot$. 


\begin{thebibliography}{1}

\bibitem{barr79}
Michael Barr.
\newblock {\em *-Autonomous Categories}.
\newblock Number 752 in {L}ecture {N}otes in {M}athematics. Springer-Verlag,
  Berlin, 1979.

\bibitem{barr96}
Michael Barr.
\newblock The {C}hu construction.
\newblock {\em Theory and {A}pplications of {C}ategories}, 2:17--35, 1996.

\bibitem{Girard94lla}
Jean-Yves Girard.
\newblock Linear {L}ogic: Its {S}yntax and {S}emantics.
\newblock In J.-Y. Girard, Y.~Lafont, and L.~Regnier, editors, {\em Advances in
  Linear Logic}, pages 1--42. Cambridge University Press, 1995.
\newblock Proceedings of the Workshop on Linear Logic, Ithaca, New York, June
  1993.

\bibitem{games91}
Y.~Lafont and T.~Streicher.
\newblock Games {S}emantics for {L}inear {L}ogic.
\newblock In {\em 6th Annual IEEE Symp. on Logic in Computer Science}, pages
  43--49, Amsterdam, July 1991.

\bibitem{Pr95}
V.~R. Pratt.
\newblock The {S}tone gamut: {A} coordinatization of mathematics.
\newblock In {\em Logic in {C}omputer {S}cience}, pages 444--454. IEEE
  {C}omputer {S}ociety, 1995.

\bibitem{Pr97b}
V.~R. Pratt.
\newblock Chu spaces from the representational viewpoint.
\newblock In {\em Parikh Festschrift}. 1997.

\bibitem{Pr94}
V.~R. Pratt.
\newblock Types as {P}rocesses, via {C}hu spaces.
\newblock In {\em EXRESS'97 Proceedings}. 1997.

\end{thebibliography}
\end{document}